\RequirePackage{ifpdf}
\ifpdf 
\documentclass[pdftex]{sigma}
\else
\documentclass{sigma}
\fi

\begin{document}
\allowdisplaybreaks

\renewcommand{\PaperNumber}{012}

\def\on#1#2{\mathop{\vbox{\ialign{##\crcr\noalign{\kern2pt}
$\scriptstyle{#2}$\crcr\noalign{\kern2pt\nointerlineskip}
\kern-2pt$\hfil\displaystyle{#1}\hfil$\crcr}}}\limits}

\FirstPageHeading

\ShortArticleName{On Classical $r$-Matrix for the Kowalevski
Gyrostat on $so(4)$}

\ArticleName{On Classical $\boldsymbol{r}$-Matrix for the
Kowalevski Gyrostat\\ on $\boldsymbol{so(4)}$}

\Author{Igor V. KOMAROV and Andrey V. TSIGANOV}

\AuthorNameForHeading{I.V. Komarov and A.V. Tsiganov}

\Address{V.A. Fock Institute of Physics, St.~Petersburg State
University, St.~Petersburg, Russia}

\Email{\href{mailto:igor.komarov@pobox.phys.spbu.ru}{igor.komarov@pobox.phys.spbu.ru},
\href{mailto:tsiganov@mph.phys.spbu.ru}{tsiganov@mph.phys.spbu.ru}}

\ArticleDates{Received November 18, 2005, in final form January
19, 2006; Published online January 24, 2006}

\Abstract{We present the trigonometric Lax matrix and classical
$r$-matrix for the Ko\-wa\-levski gyrostat on $so(4)$ algebra by
using the auxiliary matrix algebras $so(3,2)$ or $sp(4)$.}

\Keywords{Kowalevski top; Lax matrices; classical $r$-matrix}

\Classification{37J35; 70E40; 70G65}

\section{Introduction}
The classical $r$-matrix structure is an important tool for
investigating integrable systems. It encodes the Hamiltonian
structure of the Lax equation, provides the involution of
integrals of motion and gives a natural framework for quantizing
integrable systems. The aim of this paper is severalfold. First,
we present  formulae  for the classical $r$-matrices of the
Kowalevski gyrostat on Lie algebra $so(4)$, derived in the
framework of the Hamiltonian reduction. In the process we shall
get new form of its $5\times 5$ Lax matrix  and discuss the
properties of the $r$-matrices. Finally, we get the $4\times 4$
Lax matrix on the auxiliary $sp(4)$ algebra.

Remind, the Kowalevski top is the third integrable case of motion
of rigid body rotating in a constant homogeneous
field~\cite{kow89}. This is an integrable system on the orbits of
the Euclidean Lie algebra $e(3)$ with a quadratic and a quartic in
angular momenta integrals of motion.

The Kowalevski top can be generalized in several directions. We
can change either initial phase space or  the form of the Hamilton
function. In this paper we consider the Kowalevski gyrostat with
the Hamiltonian
\begin{gather} \label{Ham}
H=J_1^2+J_2^2+2J_3^2+2\rho J_3+2y_1,\qquad \rho \in\mathbb R,
\end{gather}
 on a generic orbit of the $so(4)$
Lie algebra with the Poisson brackets
\begin{gather}\label{bundle}
\bigl\{J_i,J_j\bigr\}=\varepsilon_{ijk}J_k, \qquad
\bigl\{J_i,y_j\bigr\}=\varepsilon_{ijk}y_k ,\qquad
\bigl\{y_i,y_j\bigr\}=\varkappa^2\varepsilon_{ijk}J_k,
\end{gather}
where $\varepsilon_{ijk}$ is the totally skew-symmetric tensor and
$\varkappa \in \mathbb C$ (see~\cite{kst03} for references).
Fixing values~$a$ and $b$ of the Casimir functions
\begin{gather}\label{center}
A=\sum_{i=1}^3 y_i^2 + \varkappa^2\sum_{i=1}^3 J_i^2, \qquad
B=\sum_{i=1}^3 x_iJ_i
\end{gather}
one gets a four-dimensional orbit of  $so(4)$
\[
{\cal O}_{ab}:\quad \{ y, \, J\,:\, A=a,\
 B=b\},
\]
which is a reduced phase space for the deformed Kowalevski top.

Because physical quantities $y$, $J$ in (\ref{Ham}) should be
real, $\varkappa ^2$ must be real too and   algebra (\ref{bundle})
is reduced to its two real forms  $so(4,\mathbb R)$ or
$so(3,1,\mathbb R)$ for positive and negative $\varkappa ^2$
respectively and to $e(3)$ for $\varkappa =0$.

The Hamilton function (\ref{Ham}) is fixed up to canonical
transformations. For instance, the  brac\-kets~(\ref{bundle}) are
invariant with respect to scale transformation $y_i\to cy_i$ and
$\varkappa \to c\varkappa $ that allows  to include scaling
parameter $c$ into the Hamiltonian, i.e.\ to change $y_1$ by
$cy_1$. Some other transformations are discussed in~\cite{kst03}.

Below we identify Lie algebra $\mathfrak g$ with its dual
$\mathfrak g^*$ by using invariant inner product and notation
$\mathfrak g^*$ is used both for the dual Lie algebra and  for the
corresponding Poisson manifold.

\section{The Kowalevski gyrostat: some known results}
The Lax matrices  for the Kowalevski gyrostat  was found in
\cite{RS} and \cite{kst03} at $\varkappa =0$ and $\varkappa \neq
0$ respectively. The corresponding classical $r$-matrices have
been constructed in \cite{mar98} and \cite{ts04}. In these papers
different definitions of the classical $r$-matrix \cite{rs03,bd82}
were used, which we briefly discuss below.

\subsection{The Lax matrices}

By definition the Lax matrices $L$ and $M$ satisfy  the Lax
equation
\begin{gather}\label{Lax-Eq}
\frac{d}{dt}{L(\lambda)}\equiv\bigl\{H,L(\lambda)\bigr\}=[M(\lambda),
L(\lambda)]\,
\end{gather}
with respect to evolution determined by Hamiltonian $H$. Usually
the matrices $L$ and $M$ take values in some auxiliary algebra
$\mathfrak g$ (or in its representation), whereas entries of $L$
and $M$ are functions on the phase space of a given integrable
system depending on spectral parameter $\lambda$.

For $\varkappa  =0$  the Lax matrices  for the Kowalevski gyrostat
on $e(3)$ algebra were found   by Reyman and Semenov-Tian-Shansky
\cite{RS}
\begin{gather}\label{lax-e3}
L_0(\lambda)=\left(\begin{array}{rrrrr}0 & J_3 & -J_2 &
\lambda-\dfrac{y_1}{\lambda} & 0 \vspace{2mm}\\  -J_3 & 0 & J_1 &
-\dfrac{y_2}{\lambda} & \lambda \vspace{2mm}\\   J_2 & -J_1 & 0 &
-\dfrac{y_3}{\lambda} & 0 \vspace{2mm}\\
\lambda-\dfrac{y_1}{\lambda} &
-\dfrac{y_2}{\lambda} & -\dfrac{y_3}{\lambda} & 0 & -J_3-\rho  \vspace{2mm}\\
0 & \lambda & 0 & J_3+\rho & 0
\end{array}\right)
\end{gather}
and
\begin{gather} \label{M-Kow}
M_0(\lambda)=2 \left(\begin{array}{ccccc} 0& -2J_3& J_2& -\lambda&
0 \vspace{1mm}\\
2 J_3& 0& -J_1& 0&-\lambda\vspace{1mm}\\ -J_2&J_1& 0&0&0
\vspace{1mm}\\
-\lambda&0& 0& 0&0
\vspace{1mm}\\
0&-\lambda&0& 0& 0\end{array}\right).
\end{gather}
These matrices belong to the twisted loop algebra $\mathfrak
g_\lambda$ based on the auxiliary Lie algebra $\mathfrak
g=so(3,2)$ in fundamental representation. We have to underline
that the phase space of the Kowalevski gyrostat and the auxiliary
space of these Lax matrices are essentially \textit{different}.

Remind the  auxiliary Lie algebra $so(3,2)$ may be defined by all the
$5\times 5$ matrices satisfying
\[
X^T=-JXJ,
\]
where $J=\mbox{\rm diag}\,(1,1,1,-1,-1)$,  and $T$ stands for
matrix transposition. The Cartan involution on $\mathfrak
g=so(3,2)$ is given by
\[
\sigma X=-X^T
\]
and $\mathfrak g=\mathfrak f+\mathfrak p$ is the corresponding
Cartan decomposition where $\mathfrak f=so(3)\oplus so(2)$ is the
maximal compact subalgebra of $so(3,2)$. The pairing between
$\mathfrak g$ and $\mathfrak g^*$ is given by invariant inner
product
\begin{gather}
\label{pair1} (X,Y)=-\frac12\,\mbox{\rm tr}\,XY
\end{gather}
that is positively definite on $\mathfrak f$.

We extend the involution $\sigma$ to the loop algebra $\mathfrak
g_\lambda$  by setting $(\sigma X)(\lambda)=\sigma (X(-\lambda))$.
By definition, the twisted loop algebra $\mathfrak g_\lambda$
consists of matrices $X(\lambda)$ such that
\begin{gather}\label{X-pq}
X(\lambda)=-X^T(-\lambda).
\end{gather}
The pairing between $\mathfrak g_\lambda$ and $\mathfrak
g_\lambda^*$ is given by
\begin{gather}\label{pair2}
\langle X,Y\rangle=\mbox{\rm Res }\lambda^{-1}(X,Y).
\end{gather}

At $\varkappa \neq 0$  the  Lax matrices for the Kowalevski
gyrostat on $so(4)$ were originally found  in \cite{kst03} as a
deformation of the  matrices $L_0(\lambda)$ and $M_0(\lambda)$
\begin{gather}
\label{old-lax} L=Y_{c}\cdot L_0,\qquad M=M_0\cdot Y^{-1}_c,\qquad
Y_{c}=\mathrm{diag}\left(1,1,1,\frac{\lambda^2}{\lambda^2-\varkappa
^2},1\right).
\end{gather}
Algebraic nature of the matrix $L(\lambda)$ (\ref{old-lax}) is
appeared to be mysterious, because the diagonal matrix $Y_c$ does
not belong to the fundamental representation of the auxiliary
$so(3,2)$ algebra, hence matrices (\ref{old-lax}) do not belong to
the Reyman--Semenov-Tian-Shansky scheme \cite{rs03}.

In the next section we prove that the Lax matrix $L(\lambda)$ at
$\varkappa \neq 0$ is a trigonometric deformation of the rational
Lax matrix $L_0(\lambda)$ on the same auxiliary space.

\subsection[Classical $r$-matrix: operator notations]{Classical $\boldsymbol{r}$-matrix: operator notations}

The classical $r$-matrix is a linear operator ${\mathbf r}\in
\mathrm{End}\,\mathfrak g$ that determines second Lie
bracket on $\mathfrak g$ by the rule
\[
[X,Y]_r=[{\mathbf r} X,Y]+[X,{\mathbf r} Y].
\]
The operator ${\mathbf r}$ is a classical $r$-matrix for a
given integrable system, if the corresponding equations of motion
with respect to the $r$-brackets have the Lax form (\ref{Lax-Eq})
and the second Lax matrix $M$ is given by
\begin{gather*}
M=\dfrac12\,{\mathbf r}\bigl(d H\bigr).
\end{gather*}
In the most common cases  ${\mathbf r}$ is a skew-symmetric
operator such that
\begin{gather}\label{r-mat}
{\mathbf r}=P_+-P_-,
\end{gather}
where $P_\pm$ are projection operators onto complementary
subalgebras $\mathfrak g_\pm$ of $\mathfrak g$. In this case there
exists a complete classification theory. All details may be found
in the book~\cite{rs03} and references therein.

Marchall  \cite{mar98}  has  shown that the  Lax matrices
(\ref{lax-e3}) for the Kowalevski gyrostat on $e(3)$ may be
obtained by direct application of this $r$-matrix approach. Let us
introduce the standard decomposition of any element $X\in\mathfrak
g_\lambda$
\begin{gather}\label{decom-g}
X(\lambda)= X_+(\lambda) + X_0+ X_-(\lambda),
\end{gather}
where $X_+(\lambda)$ is a Taylor series in $\lambda$, $X_0$ is an
independent of $\lambda$ and $X_-(\lambda)$ is a series in
$\lambda^{-1}$. If $P_\pm$ and $P_0$ are the projection operators
onto $\mathfrak g_\lambda$ parallel to the complementary
subalgebras (\ref{decom-g}), the operator
\begin{gather}\label{r-marsh}
{\mathbf r}=P_-+\varrho\circ P_0-P_+.
\end{gather}
defines the second Lie structure on  $\mathfrak g_\lambda$.
According to~\cite{mar98} the $r$-matrix (\ref{r-marsh}) is the
classical $r$-matrix for the Kowalevski gyrostat. In the standard
case (\ref{r-mat}) operator $\varrho$ is identity, however for the
Kowalevski gyrostat  $\varrho$ is a difference of projectors in
the base $\mathfrak g=so(3,2)$ (see details in~\cite{mar98}).

\subsection[Classical $r$-matrix: tensor notations]{Classical $\boldsymbol{r}$-matrix: tensor notations}

Another definition of the classical $r$-matrix is more familiar in
the inverse scattering method \cite{rs03,bv90,bd82}. According
to \cite{bv90}, the commutativity of the spectral invariant of the
matrix $L(\lambda)$ is equivalent to existence of a classical
$r$-matrix $r_{12}(\lambda,\mu)$ such that  the Poisson brackets
between the entries of $L(\lambda)$ may be rewritten in the
following commutator form
\begin{gather}\label{rpoi}
\big\{\,{\on{L}{1}}(\lambda),\,{\on{L}{2}}(\mu)\,\big\}=
\big[r_{12}(\lambda,\mu),\,{\on{L}{1}}(\lambda)\big]
-\big[r_{21}(\lambda,\mu),\,{\on{L}{2}}(\mu)\,\big].
\end{gather}
Here
\[{\on{L}{1}}(\lambda)= L(\lambda)\otimes 1,\qquad
{\on{L}{2}}(\mu)=1\otimes L(\mu),\qquad
r_{21}(\lambda,\mu)=\Pi\,r_{12}(\mu,\lambda)\,\Pi,\] and $\Pi$ is
a permutation operator $\Pi X\otimes Y=Y\otimes X\Pi$ for any
numerical matrices $X$, $Y$.

For a given  Lax matrix $L(\lambda)$, $r$-matrices are far from
being uniquely defined. The possible ambiguities are discussed in
\cite{rs03,bv90,bd82}.

If the Lax matrix takes values in some Lie algebra $\mathfrak g$
(or in its representation), the $r$-matrix takes values in
$\mathfrak g\times\mathfrak g$ or its corresponding
representation. The matrices $r_{12}$, $r_{21}$ may be identified
with  kernels of the operators ${\mathbf r}\in \mathrm{End}\,
\mathfrak g$ and ${\mathbf r}^*\in \mathrm{End}\, \mathfrak
g^*$ respectively, using pairing between $\mathfrak g$ and
$\mathfrak g^*$ (see discussion in~\cite{rs03}).

Generally speaking, the matrix $r_{12}(\lambda,\mu)$ is a function
of dynamical variables \cite{bd82,skl94}. In the most extensively
studied case of  purely numeric $r$-matrices it  satisfies the
classical Yang--Baxter equation
\begin{gather}\label{cybe}
\left[r_{12}(\lambda,\mu),r_{13}(\lambda,\nu)+r_{23}(\mu,\nu)\right]
-\left[r_{13}(\lambda,\nu),r_{32}(\nu,\mu)\right]=0,
\end{gather}
which ensures the Jacobi identity for the Poisson brackets
(\ref{rpoi}). If $r_{12}(\lambda,\mu)$ is a unitary numeric matrix
depending on the difference of the spectral parameters
$z=\lambda-\mu$, there exists a profound algebraic theory, which
allows to classify $r$-matrices in various
families~\cite{bd82,rs03}.

For the Kowalevski gyrostat  the classical $r$-matrix
$r_{12}(\lambda,\mu)$ entering (\ref{rpoi})  has been constructed
in \cite{ts04} by using the auxiliary Lie algebra $\mathfrak
g=so(3,2)$ in fundamental representation. The generating set of
this auxiliary space consists of one antisymmetric matrix
\[S_4=\left(\begin{array}{ccccc} 0&
0 & 0 & 0 & 0 \\ 0 & 0 & 0 & 0 & 0 \\ 0 & 0 & 0 & 0 & 0 \\ 0 & 0 & 0 & 0 & -1 \\
0 & 0 & 0 & 1 & 0
\end{array}\right)\]
and three symmetric matrices
\[
 Z_1=\left(\begin{array}{ccccc} 0 & 0 & 0 & 1 & 0 \\ 0 & 0 & 0 & 0 & 0 \\
0 & 0 & 0 & 0 & 0 \\ 1 & 0 & 0 & 0 & 0 \\ 0 & 0 & 0 & 0 &
0\end{array}\right) ,\quad
 Z_2= \left(\begin{array}{ccccc} 0 & 0 & 0 & 0 & 0 \\
 0 & 0 & 0 & 1 & 0 \\  0 & 0 & 0 & 0 & 0 \\  0 & 1 & 0 & 0 & 0 \\ 0
& 0 & 0 & 0 & 0\end{array}\right),\quad
Z_3=\left(\begin{array}{ccccc} 0 & 0 & 0 & 0 & 0 \\ 0 & 0 & 0 & 0
& 0 \\  0 & 0 & 0 & 1 & 0 \\  0 & 0 & 1 & 0 & 0 \\  0 & 0 & 0 & 0
& 0\end{array}\right),
\]
which are the generators of the $so(3,2)$ algebra. Other
generators are three symmetric matrices
\begin{gather}\label{h-matr}
H_i=[S_4,Z_i]\equiv S_4Z_i-Z_iS_4, \qquad i=1,2,3.\end{gather} and
three antisymmetric matrices
\begin{gather}\label{s-matr}
S_1=[Z_2,Z_3],\qquad S_2=[Z_3,Z_1],\qquad S_3=[Z_1,Z_2] .
\end{gather}
These matrices are orthogonal with respect to the form of trace
(\ref{pair1}). Four matrices $S_k$ form maximal compact subalgebra
$\mathfrak f=so(3)\oplus so(2)$ of $so(3,2)$ and their norm is
$1$, whereas six matrices $Z_i$ and $H_i$ belong to the
complementary subspace $\mathfrak p$ in the Cartan decomposition
$\mathfrak g=\mathfrak f+\mathfrak p$ and their norms are $-1$.
Operators
\[
P_{\mathfrak f}=\sum_{k=1}^4 S_k\otimes S_k,\qquad\mbox{\rm
and}\qquad P_{\mathfrak p}=\sum_{i=1}^3 \bigl(H_i\otimes
H_i+Z_i\otimes Z_i\bigr)
\]
are projectors onto the orthogonal subspaces $\mathfrak f$ and
$\mathfrak p$ respectively.

In this basis the Lax matrix $L_0(\lambda)$ (\ref{lax-e3}) for the
Kowalevski gyrostat on $e(3)$ reads as
\[
L_0=\lambda(Z_1+H_2)+\sum_{i=1}^3
\left(J_iS_i-\lambda^{-1}\,x_i\,Z_i\right)+(J_3+\rho)S_4.
\]
According to~\cite{ts04} the corresponding $r$-matrix is equal to
\begin{gather}
r_{12}(\lambda,\mu)=\dfrac{\lambda\mu}{\lambda^2-\mu^2}\,P_{\mathfrak
p}-\dfrac{\mu^2}{\lambda^2-\mu^2}\,P_{\mathfrak
f}+\bigl(S_3-S_4\bigr)\otimes S_4\nonumber\\
\phantom{r_{12}(\lambda,\mu)}{}
=\dfrac{\lambda\mu}{\lambda^2-\mu^2}\sum_{i=1}^3\bigl(H_i\otimes
H_i+Z_i\otimes Z_i\bigr)-\dfrac{\mu^2}{\lambda^2-\mu^2}
\sum_{k=1}^4 S_k\otimes S_k+\bigl(S_3-S_4\bigr)\otimes
S_4.\!\!\!\label{r-rat}
\end{gather}
We can say that this matrix $r_{12}(\lambda,\mu)$ is a
\textit{specification} of the operator ${\mathbf r}$
(\ref{r-marsh}) with respect to canonical pairing
(\ref{pair1})--(\ref{pair2}).

In  applications to integrable models,  see
e.g.~\cite{rs03,bv90,bd82}, the solutions of the Yang--Baxter
equation (\ref{cybe}) given by  unitary numeric matrices,
satisfying the relation
\begin{gather*}
r_{12}(\lambda,\mu)=-r_{21}(\mu,\lambda)
\end{gather*}
and depending on the difference $\lambda-\mu$,  had been studied
most extensively.

In our case the matrix $r_{12}(\lambda,\mu)$ (\ref{r-rat}) is
appeared to be purely numeric matrix, which depends on the ratio
$\lambda/\mu$ only. It allows us to change the spectral parameters
$\lambda=e^{u_1}$ and $\mu=e^{u_2}$ and rewrite this $r$-matrix in
the following form
\begin{gather*}
r_{12}(z)=\dfrac1{2\sinh(z)} P_{\mathfrak p}-\dfrac{1}{2\sinh
z(\cosh z+\sinh z)} P_{\mathfrak f}+ (S_3-S_4)\otimes S_4,
\end{gather*}
depending on one parameter $z=u_1-u_2$ via trigonometric
functions. Therefore, the classical $r$-matrix for the Kowalevski
gyrostat on $e(3)$ should be considered as \textit{trigonometric}
$r$-matrix according to generally accepted classification
\cite{bd82,rs03}.

At the same time it is natural to keep  initial rational
parameters $\lambda$, $\mu$ in the Lax matrix. Similar properties
holds for the periodic Toda chain, for which
 $N \times N$ Lax matrix depends rationally on spectral parameters, while the corresponding
 $r$-matrix is trigonometric.

We have to underline that in contrast with usual cases this
$r$-matrix is non-unitary. Moreover, it has a term
$(S_3-S_4)\otimes S_4$, which is independent on spectral
parameters and, therefore, the inequality $r_{12}(z)\ne
-r_{12}(-z)$ takes place.

In order to understand the nature of these items we  recall that
the Lax matrix $L_0(\lambda)$ has been derived in the framework of
the Hamiltonian reduction of the $so(3,2)$ top for which phase
space coincides with the auxiliary space. The corresponding
classical $r$-matrix, calculated in~\cite{ts04}
\[
r_{12}^{\mathrm{so(3,2)}}(z)=\dfrac1{2\sinh(z)}\left(P_{\mathfrak
p}-\dfrac{1}{\cosh z+\sinh z} P_{\mathfrak f}\right)
\]
is a  trigonometric $r$-matrix associated with the $so(3,2)$ Lie
algebra~\cite{bd82}. So, the constant term  $(S_3-S_4)\otimes S_4$
in (\ref{r-rat}) is an immediate result of the Hamilton reduction,
which changes the phase space of our integrable system.

We recall that classical $r$-matrices $r_{12}(\lambda,\mu)$ are
called regular solutions to the Yang--Baxter equation~(\ref{cybe})
if they pass through the unity at some $\lambda$ and $\mu$. In our
case we have the following counterpart of this property of
regularity
\[
\left.\mbox{\rm res}\,r_{12}(z)\right|_{z=0}=\dfrac12
\bigl(P_{\mathfrak p}-P_{\mathfrak f}\bigr).
\]

\section[Classical $r$-matrix for Kowalevski gyrostat on $so(4)$]{Classical
 $\boldsymbol{r}$-matrix for Kowalevski gyrostat on $\boldsymbol{so(4)}$}

Now let us consider  Lax matrix $L(\lambda)$ (\ref{old-lax}) for
the Kowalevski gyrostat on $so(4)$ algebra. After transformation
$L(\lambda)\to  \cos\phi \,Y_{c}^{-1/2}L(\lambda)Y_{c}^{1/2}$  of
the Lax matrix $L(\lambda)$ (\ref{old-lax}) and change of the
spectral parameter $\lambda=\varkappa /\sin\phi$ one gets a
trigonometric Lax matrix on the auxiliary $so(3,2)$ algebra
\begin{gather}
\label{lax-o4} L=\dfrac{\varkappa}{\sin\phi}\bigl(Z_1+\cos\phi
H_2\bigr)+\sum_{i=1}^3 \left(\cos\phi\,J_i
S_i-\varkappa^{-1}\sin\phi\,y_iZ_i\right)+(J_3+\rho)S_4
\end{gather}
or
\begin{gather}
\label{lax-55} L=\left(\begin{array}{rrrrr}0 & \cos\phi\,J_3 &
-\cos\phi\,J_2 & \dfrac{\varkappa}{\sin\phi}-
{\dfrac{\sin\phi}{\varkappa}\,y_1}  & 0
\vspace{2mm}\\
-\cos\phi\,J_3 & 0 & \cos\phi\,J_1 & -
{\dfrac{\sin\phi}{\varkappa}\,y_2}  &
\dfrac{\varkappa\cos\phi}{\sin\phi}
\vspace{2mm}\\
 \cos\phi\,J_2 &
-\cos\phi\,J_1 & 0 & - {\dfrac{\sin\phi}{\varkappa}\,y_3}  & 0
\vspace{2mm}\\
\dfrac{\varkappa}{\sin\phi}- {\dfrac{\sin\phi}{\varkappa}\,y_1}  &
- {\dfrac{\sin\phi}{\varkappa}\,y_2} & -
{\dfrac{\sin\phi}{\varkappa}\,y_3} & 0 & -J_3-\rho
\vspace{2mm}\\
0 & \dfrac{\varkappa\cos\phi}{\sin\phi} & 0 & J_3+\rho & 0
\end{array}\right).
\end{gather}
In order to consider the real forms $so(4,\mathbb R)$ or
$so(3,1,\mathbb R)$ we have to use trigonometric or hyperbolic
functions for positive and negative $\varkappa ^2$, respectively.

If we put $\phi=\varkappa\lambda^{-1}$ and take the limit
$\varkappa \to 0 $ we find the rational Lax matrix $L_0(\lambda)$
(\ref{lax-e3}) for the Kowalevski gyrostat on $e(3)$.

The Lax matrices $L(\phi)$ and $L_0(\lambda)$ are invariant with
respect to the following involutions
\begin{gather}\label{new_inv}
L(\phi)\to -L^T(-\phi)\qquad\mbox{\rm and}\qquad L_0(\lambda)\to
-L_0^T(-\lambda),
\end{gather}
that are compatible with the Cartan involution $\sigma$. This
simple observation shows  that for the Kowalevski  $so(4)$
gyrostat the Reyman--Semenov-Tian-Shansky scheme~\cite{rs03}
should be extended from rational to trigonometric case.

One can prove that the trigonometric Lax matrix $L(\phi)$
(\ref{lax-o4}) satisfies relation
\begin{gather*}
\big\{\,{\on{L}{1}}(\phi), \,{\on{L}{2}}(\theta)\,\big\}=
\big[\,r_{12}(\phi,\theta),\, {\on{L}{1}}(\phi)\,\big]
-\big[r_{21}(\phi,\theta),\,{\on{L}{2}}(\theta)\,\big].
\end{gather*}
with the following $r$-matrix
\begin{gather}
r_{12}(\phi,\theta)=
\frac{\sin\phi\sin\theta}{\cos^2\phi-\cos^2\theta}\sum_{i=1}^3\bigl(
\cos\theta\, H_i\otimes H_i+\cos\phi\, Z_i\otimes
Z_i\bigr)\nonumber\\
\phantom{r_{12}(\phi,\theta)=}{}
-\frac{\sin^2\phi}{\cos^2\phi-\cos^2\theta}\sum_{k=1}^4
\cos\theta\,S_k\otimes S_k+
\left(S_3-\dfrac{\cos\phi\cos\theta+1}{\cos\phi+\cos\theta}\,S_4\right)\otimes
S_4.\label{r-trig}
\end{gather}
If we put $\phi=\varkappa\lambda^{-1}$, $\theta=\varkappa\mu^{-1}$
and take the limit $\varkappa \to 0$ we get classical $r$-matrix
for the Kowalevski gyrostat on $e(3)$ algebra (\ref{r-rat}). As
above the matrix $r_{12}(\phi,\theta)$ satisfies the Yang--Baxter
equation~(\ref{cybe}) and it has the same analog of the property
of regularity
\[
\left.\mbox{\rm
res}\,r_{12}(\phi,\theta)\right|_{\phi=\theta}\simeq
\bigl(P_{\mathfrak p}-P_{\mathfrak f}\bigr).
\]
In contrast with $r$-matrix (\ref{r-rat}) for  the Kowalevski
gyrostat on $e(3)$ we can not rewrite this
$r$-mat\-rix~(\ref{r-trig}) as a function depending on the
difference of the spectral parameters only. We suggest that it may
be  possible to present it in terms of elliptic functions  of one
spectral parameters after a proper similarity transformation and
reparametrization.

The well known isomorphism between $so(3,2)$ and $sp(4)$ algebras
allows us to consider $4\times 4$ Lax matrix instead of $5\times
5$ matrix (\ref{lax-55}). The generating set $Z_1$, $Z_2$, $Z_3$
and $S_4$ may be represented by different $4\times 4$ real or
complex matrices, for instance,
\[ s_4=\frac{i}{2}\left(\begin{array}{rrrr} -1& 0 & 0 & 0 \\
0 & 1 & 0 & 0 \\ 0 & 0 & -1 & 0 \\ 0 & 0 & 0 & 1
\end{array}\right),\]
and
\[
z_1= \frac{i}{2}\!\left(\begin{array}{rrrr} 0 \- & 1 & 0 & 0 \\ -1 & 0 & 0 & 0 \\
 0 & 0 & 0 & -1 \\0 & 0 & 1 & 0 \end{array}\right),\quad
 z_2=\frac{1}{2}\!\left(\begin{array}{rrrr} 0 & 1 & 0 & 0 \\
 1 & 0 & 0 & 0 \\  0 & 0 & 0 & 1 \\  0 & 0 & 1 & 0 \end{array}\right),\quad
  z_3=\frac{i}2\!\left(
\begin{array}{rrrr} 0 & 0 &  0 & 1 \\  0 & 0 & -1 & 0 \\  0 &
1 & 0 & 0 \\  -1 & 0 & 0 & 0 \end{array}\right).
\]
Other $sp(4)$ generators are constructed by
(\ref{h-matr})--(\ref{s-matr}). These matrices are orthogonal with
respect to the form of trace (\ref{pair1}). Norm of matrices $s_k$
is $1/2$, whereas six matrices $z_i$ and $h_i$ have norm~$-1/2$.

In this basis the $4\times 4$ Lax matrix for the Kowalevski
gyrostat on $so(4)$ reads
\begin{gather}
\label{L4_tr} L^{(4)}(\phi)= \left(\begin{smallmatrix}
(1-\cos\phi) J_3+\rho& \frac{\sin\phi}{\varkappa }
y_--\frac{\varkappa \sin\phi}{\cos\phi+1}& \cos\phi J_-&
\frac{\sin\phi}{\varkappa }y_3\vspace{1mm}\\
-\frac{\sin\phi}{\varkappa }y_++\frac{\varkappa
\sin\phi}{\cos\phi+1}& (\cos\phi-1)J_3-\rho&
-\frac{\sin\phi}{\varkappa }y_3& -\cos\phi J_+\vspace{1mm}\\
\cos\phi J_+&\frac{\sin\phi}{\varkappa }y_3& (\cos\phi+1)J_3+\rho&
-\frac{\sin\phi}{\varkappa }y_++\frac{\varkappa
(\cos\phi+1)}{\sin\phi}\vspace{1mm}\\
-\frac{\sin\phi}{\varkappa }y_3& -\cos\phi J_-&
\frac{\sin\phi}{\varkappa }y_--\frac{\varkappa
(\cos\phi+1)}{\sin\phi}& -(\cos\phi+1)J_3-\rho
\end{smallmatrix}
\right).
\end{gather}
Here $J_\pm=J_1\pm iJ_2$ and $y_\pm=y_1\pm iy_2$.

If we put $\phi=\varkappa\lambda^{-1}$ and take the limit
$\varkappa \to 0 $ we find the rational Lax matrix for the
Kowalevski gyrostat on $e(3)$
\begin{gather*}
L_0^{(4)}(\lambda)= \left(
\begin{array}{rrrr}
\rho& \dfrac{y_-}{\lambda}&  J_-& \dfrac{y_3}{\lambda}\vspace{2mm}\\
-\dfrac{y_+}{\lambda}& -\rho&
-\dfrac{y_3}{\lambda}& -J_+\vspace{2mm}\\
 J_+&\dfrac{y_3}{\lambda}& 2J_3+\rho&
 -\dfrac{y_+}{\lambda}+2\lambda\vspace{2mm}\\
-\dfrac{y_3}{\lambda}& -J_-& \dfrac{y_-}{\lambda}-2\lambda&
-2J_3-\rho
\end{array}
\right).
\end{gather*}
According to \cite{bk88}, this matrix has a mysterious property.
Namely, it contains the $3\times 3$ Lax matrix ${\widehat
L}(\lambda)$ for the Goryachev--Chaplygin gyrostat on $e(3)$
algebra as its $(1,1)$-minor. Remind that the Goryachev--Chaplygin
gyrostat with Hamiltonian
\[
\hat H=J_1^2+J_2^2+(2J_3+\rho)^2+4x_1
\]
is an integrable system at the zero value $b=0$ of the Casimir
element $B$~(\ref{center}) only. Similar pro\-perty obeys
$(2,2)$-minor of $L_0^{(4)}(\lambda)$. These properties strongly
depend on the chosen basis~$s_{i}$,~$z_{i}$ and $h_{i}$.

It is easy to prove that $(1,1)$-minor of the trigonometric Lax
matrix (\ref{L4_tr}) cannot be a Lax matrix for any integrable
system. It is compatible with the known fact that the
Goryachev--Chaplygin gyrostat on $e(3)$ cannot be naturally lifted
to $so(4)$ algebra.

\section{Conclusion}

There are few Lax matrices obtained for  deformations of known
integrable systems from their undeformed counterpart in the form
(\ref{old-lax})  (see \cite{kst03,ts04} and references within).
The important question in construction of these matrices by the
Ansatz $L=Y_{c}\cdot L_0$~(\ref{old-lax}) is a choice of a proper
matrix $Y_{c}$ for a given rational matrix $L_0(\lambda)$. In all
known cases  this transformation destroys the original auxiliary
algebra, because the corresponding matrices  $Y_{c}$ do not belong
it.

In this note we show that if one takes a Lax matrix of the
Kowalevski $so(4)$ gyrostat in  the symmetric form
$L=Y_{c}^{1/2}\cdot L_0\cdot Y_{c}^{1/2} $ and makes a
trigonometric change of spectral parameter it restores the
original auxiliary $so(3,2)$ algebra and new $L$ respects the
trigonometric  current involution (\ref{new_inv}). It means that
deformation of the physical space from the orbits of  $e(3)$ to
that of $so(4)$ algebra is naturally  related with transition from
rational to trigonometric parametrization of the auxiliary
current algebra.

We calculated explicitly the corresponding  $r$-matrices and
demonstrated that constant terms in them is due to the Hamiltonian
reduction.

The  classical $r$-matrix (\ref{r-trig}) for the $so(4)$ gyrostat
is numeric  and the corresponding Lax matrix $L(\phi)$
(\ref{lax-o4}) does not contain ordering problem in quantum
mechanics. Hence equation (\ref{rpoi}) holds true in quantum case
both for the Lax matrices (\ref{lax-o4}) and (\ref{L4_tr}).

\subsection*{Acknowledgements}
The authors thank E.~Sklyanin and N.~Reshetikhin for very useful
conversations on the subject of this paper.
 I.V.K.~wishes to thank the London Mathematical Society for support his visit to England
 and V.B.~Kuznetsov for hospitality at the  University of Leeds.

\LastPageEnding

\end{document}